\documentclass[12pt]{article}
\usepackage{amsmath}
\usepackage{epsf}
\usepackage{epsfig}
\usepackage{amssymb}
\usepackage{graphicx}
\usepackage{latexsym}
\newcommand{\be}{\begin{equation}}
\newcommand{\ee}{\end{equation}}
\newcommand{\bear}{\begin{eqnarray}}
\newcommand{\eear}{\end{eqnarray}}

\newsavebox{\LSIM}
\sbox{\LSIM}{\raisebox{-1ex}{$\ \stackrel{\textstyle<}{\sim}\ $}}
\newcommand{\lsim}{\usebox{\LSIM}}
\newsavebox{\GSIM}
\sbox{\GSIM}{\raisebox{-1ex}{$\ \stackrel{\textstyle>}{\sim}\ $}}

\begin{document}
\begin{titlepage}
\begin{flushright}
CERN-PH-TH/2006-064\\
BI-TP 2006/10\\
hep-ph/0604159
\end{flushright}
$\mbox{ }$
\vspace{.1cm}
\begin{center}
\vspace{.5cm}
{\bf\Large Top transport in electroweak baryogenesis}\\[.3cm]
\vspace{1cm}
Lars Fromme$^{a,}$\footnote{fromme@physik.uni-bielefeld.de}
and 
Stephan J.~Huber$^{b,}$\footnote{stephan.huber@cern.ch}
 
\vspace{1cm} {\em  
$^a$Fakult\"at f\"ur Physik, Universit\"at Bielefeld, D-33615 Bielefeld, Germany}\\[.2cm] 
{\em $^b$Theory Division, CERN, CH-1211 Geneva 23, Switzerland} 
\end{center}
\bigskip\noindent
\vspace{1.cm}
\begin{abstract}
In non-supersymmetric models of electroweak baryogenesis the top quark
plays a crucial role.  Its CP-violating source term can be calculated in the WKB
approximation.
We point out how to resolve certain discrepancies between computations starting
from the Dirac equation and the Schwinger--Keldysh formalism. We also improve
on the transport equations, keeping the W-scatterings at finite rate.
We apply these results to a model with one Higgs doublet, augmented by dimension-6
operators, where our refinements lead to an increase in the baryon 
asymmetry by a factor of up to about 5.
\end{abstract}
\end{titlepage}
\section{Introduction}
Electroweak baryogenesis \cite{Kuzmin} is typically described by a set of 
transport equations, which are fueled by CP-violating source terms. The source terms
arise from the CP-violating interactions of particles in the hot plasma with
the expanding bubble walls during a first order electroweak phase transition \cite{CKN}.
By diffusion the sources move into the symmetric phase \cite{CKN1}, where 
baryon number violation is fast.
For walls much thicker than the inverse transition temperature the wall-plasma interactions
can be treated in a WKB approximation, which corresponds to an expansion
in gradients of the bubble profile. At first order in gradients a CP-violating shift
is induced in the dispersion relations of particles crossing the bubble wall \cite{JPT95}. 
A (semiclassical) force results, different for particles and antiparticles, which creates
a non-zero left-handed quark density in front of the bubble. The weak sphalerons
partly transform this left-handed quark density into a baryon asymmetry.

The WKB approach has been widely used to study electroweak baryogenesis
in various extensions of the standard model (SM) 
\cite{CJK98,CJK00a,CJK00,NMSSM, MSSM,BFHS04,F05}.  
(An alternative approach was followed in ref.~\cite{CQW}.)
In the simplest manner, the WKB dispersion relations were computed by solving
the one-particle Dirac equation to first order in gradients in the CP-violating 
bubble wall background. In a more rigorous treatment similar dispersion
relations were also derived in the Schwinger--Keldysh formalism \cite{PSW,PSW1}.

Comparing the dispersion relation of a single Dirac fermion obtained from the 
Dirac equation \cite{CJK00}
to that of the Schwinger--Keldysh formalism \cite{PSW1}, we observe that the CP-violating
part of the latter is somewhat enhanced. In this letter we show that this
mismatch disappears when the result gained from the Dirac equation is correctly
boosted to a general Lorentz frame. Thus in the case of a single Dirac fermion, 
the full Schwinger--Keldysh result can be obtained in a much simpler way.

To demonstrate the numerical significance of this effect we recompute the baryon
asymmetry in the SM augmented by dimension-6 operators \cite{BFHS04}. Using the correct
dispersion relations enhances the baryon asymmetry by a factor of up to about 2.
We also improve on the transport equations, keeping scatterings with W
bosons at a finite rate, which considerably reduces or enhances the baryon asymmetry,
depending on the wall velocity. We also show that the position dependence of
certain thermal averages in the transport equations has a substantial impact on the
baryon asymmetry. Finally, we
investigate to what extent the CP-violating source terms are influenced by 
CP-conserving perturbations in the plasma, an effect that turns out to be negligible.
In total, depending on the model parameters, 
our refinements can increase the baryon asymmetry by a factor of up to about 5.

\section{The semiclassical force}
We consider a single Dirac fermion, such as the top quark. Its mass changes as
it passes the bubble wall. Once the bubble has sufficiently grown, we can 
approximate the bubble wall by a planar profile. The profile is kink-shaped 
and characterized by a wall thickness $L_w$. The problem is most simply 
treated in the rest frame of the bubble wall. In the presence of 
CP-violation, the fermion mass term can be complex, 
i.e.~${\rm Re}({\cal M})+i\gamma^5{\rm Im}({\cal M})$, where
\be
{\cal M}=m(z)e^{i\theta(z)}
\ee
and $z$ is the coordinate perpendicular to the bubble wall. 

For a particle with momentum much larger than $L_w^{-1}$ we can solve
the Dirac equation using a WKB ansatz
\be
\Psi\sim e^{-i\omega t+i\int^z p_{cz}(z')dz'}
\ee
and expand in gradients of ${\cal M}$. Here  $p_{cz}$ is the canonical 
momentum along the $z$ direction.
To simplify the solution we have boosted to the frame where the momentum
perpendicular to the wall is zero. Since the typical momentum of a particle in the plasma
is on the order of the temperature $T$, this approach is valid for thick bubbles,
i.e.~$TL_w\gg1$. Note that at this stage the fermion
is treated as a free particle. Scatterings with particles in the plasma 
will be incorporated later on by means of the Boltzmann equation.

As shown in ref.~\cite{CJK00}  the dispersion relation is, to first order in gradients 
\be \label{disp1}
\omega=\sqrt{(p_{cz}-\alpha_{CP})^2+m^2}\mp\frac{s\theta'}{2},
\ee
with $\theta'=\partial_z\theta$,
$\alpha_{CP}=\alpha'\pm\frac{\theta'}{2}$, and $s=1~(-1)$ for $z$-spin up (down).
The upper (lower) sign corresponds to particles and antiparticles,
respectively, which this way get different dispersion relations. 
The additional phase $\alpha$ is related to an ambiguity
in the definition of the canonical momentum, when replacing
$\Psi\rightarrow e^{i\alpha(z)}\Psi$. It was the main result of refs.~\cite{CJK00a,CJK00}
that this ambiguity disappears when all quantities are expressed in terms
of the kinetic momentum rather than the canonical momentum. 

In ref.~\cite{CJK00} the dispersion relation (\ref{disp1}) was used to compute the
semiclassical force, which was then generalized to a Lorentz frame with finite momentum
parallel to the wall. The point we make in this letter is that first
eq.~(\ref{disp1}) should be boosted to the general frame and all further 
manipulations should be
carried out later on. We will demonstrate that this way the dispersion relation
of ref.~\cite{PSW1} is correctly reproduced. 

Since Lorentz invariance is not broken parallel to the wall, we simply have to replace
$\omega^2\rightarrow \omega^2+p_{x}^2+p_{y}^2$. Note that parallel to the wall we 
do not have
to distinguish between kinetic and canonical momentum, i.e.~$p_{cx,y}=p_{x,y}$.
The dispersion relation (\ref{disp1}) turns into
\be \label{disp2}
\omega=\omega_0\mp s\frac{\theta'}{2}\frac{\omega_{0z}}{\omega_0},
\ee
where
\bear
\omega_0&=&\sqrt{(p_{cz}-\alpha_{CP})^2+p_x^2+p_y^2+m^2}
\nonumber\\[.3cm]
\omega_{0z}&=&\sqrt{(p_{cz}-\alpha_{CP})^2+m^2}.
\eear
In the limit $\omega_{0}=\omega_{0z}$ we are back at the old result. In the following we
show that when written in terms of the kinetic momentum the dependence on 
$\alpha_{CP}$ still drops. 

The physical kinetic $z$-momentum is given by $p_z=\omega v_{gz}$, where $v_{gz}$,
the group velocity of the WKB wave-packet in the $z$ direction, is given by
\be \label{v_gz}
v_{gz}=\left(\frac{\partial \omega}{\partial p_{cz}}\right)_{z}
=\frac{p_{cz}-\alpha_{CP}}{\omega_0}\left(1\mp s\frac{\theta'}{2}
\frac{\omega_0^2-\omega_{0z}^2}{\omega_0^2\omega_{0z}}\right).
\ee
The kinetic momentum then is
\be \label{pz}
p_z=(p_{cz}-\alpha)\left(1\mp s\frac{\theta'}{2\omega_{0z}}\right). 
\ee
We can use this expression to replace the canonical momentum in the
dispersion relation (\ref{disp2}).
To stress the difference, we introduce a new symbol,  $E$, to
denote energy expressed in terms of the kinetic momentum. Defining
\bear
E_0&=&\sqrt{p_z^2+p_x^2+p_y^2+m^2}
\nonumber\\[.3cm]
E_{0z}&=&\sqrt{p_z^2+m^2},
\eear
we obtain, to first order in gradients
\bear\label{Ek}
E&=&E_0\pm\Delta E=
\nonumber\\[.3cm]
&=&E_0\mp s\frac{\theta'm^2}{2E_0E_{0z}}.
\eear
Notice that the ambiguity related to $\alpha_{CP}$ has disappeared. 
For the group velocity we now find 
\be \label{vk}
v_{gz}=\frac{p_z}{E_0}\left(1\pm s\frac{\theta'}{2}
\frac{m^2}{E_0^2E_{0z}}\right).
\ee

From the canonical equations of motion we can compute the force acting on the particle 
\bear
F_z=\dot{p}_z=\omega\dot{v}_{gz}&=&
\omega\left(\dot{z}\left(\frac{\partial v_{gz}}{\partial z}\right)_{p_{cz}}
+\dot{p}_{cz}\left(\frac{\partial v_{gz}}{\partial p_{cz}}\right)_{z}\right)
\nonumber\\
&=&
\omega\left(v_{gz}\left(\frac{\partial v_{gz}}{\partial z}\right)_{p_{cz}}
-\left(\frac{\partial \omega}{\partial z}\right)_{p_{cz}}\left(\frac{\partial v_{gz}}
{\partial p_{cz}}\right)_{z}\right)
\eear
where we have used the fact that $\omega$ is constant along the trajectory. Performing the
partial derivatives and replacing the canonical by the kinetic momentum, we finally
obtain
\be \label{Fk}
F_z=-\frac{(m^2)'}{2E_0}\pm s\frac{(m^2\theta')'}{2E_0E_{0z}}
\mp s\frac{\theta'm^2(m^2)'}{4E_0^3E_{0z}}.
\ee 
Thus particles and antiparticles experience a different force as they pass the bubble wall.
This CP-violating part of the force is second order in derivatives. There is also a
CP-conserving part, which is first order in derivatives.

Our expressions for the dispersion relation (\ref{Ek}), the group velocity (\ref{vk}),
and the semiclassical force (\ref{Fk}) agree with the results of ref.~\cite{PSW1},
demonstrating that for a single Dirac fermion the full Schwinger--Keldysh result can 
be obtained in a much simpler way by means of the Dirac equation. This is the
main result of this letter. 

In the special case $E_0=E_{0z}$, i.e.~when the particle has no momentum parallel
to the wall, our results agree with those of ref.~\cite{CJK00}. For a relativistic particle 
in the plasma $E_{0z}$ contains only roughly a third of the total energy. Keeping
correct track of the factors $E_{0z}$ enhances the CP-violating part of the dispersion
relation and the force term by a factor of up to about 3. For non-relativistic particles the
effect is smaller. This factor has been neglected so far 
in computations of the baryon asymmetry based on the WKB approximation of the
Dirac equation. We will demonstrate this enhancement in a numerical example in section 4.

In the next section we discuss the impact of the CP-violating force on the transport 
equations of particles in the plasma. In a chiral theory, as the SM, interactions are 
related to the chirality of a particle rather than its spin. Thus it is convenient to label 
particles in terms of helicity 
$\lambda$, which is close to chirality for relativistic particles. 
We then have to replace the spin by $s=\lambda{\rm sign}(p_z)$ in 
eqs.~(\ref{Ek}), (\ref{vk}) and (\ref{Fk}). 

\section{Transport equations}
In the derivation of the transport equations we closely follow ref.~\cite{CJK00}.
A crucial assumption made in that work is that it is the kinetic momentum
that is conserved in the scatterings of WKB particles. The equilibrium phase space
distributions should therefore also be written in terms of the kinetic momentum. In the
wall frame this reads
\be \label{feq}
f_{i}^{(\rm eq)}({\bf x},{\bf p})=\frac{1}{e^{\beta\gamma_w(E_i+v_wp_z)}\pm1}
\ee
where $\beta=1/T$ and $\gamma_w=1/\sqrt{1-v_w^2}$, and plus (minus) refers to
fermions (bosons), respectively. We model the perturbations from equilibrium caused by the 
passage of the bubble wall with a fluid-type ansatz
\be
f_{i}({\bf x},{\bf p})=\frac{1}{e^{\beta[\gamma_w(E_i+v_wp_z)-\mu_i]}\pm1}+
\delta f_{i}({\bf x},{\bf p}).
\ee
The chemical potentials $\mu_i(z)$ describe a local departure from the equilibrium 
particle density.
The perturbations $\delta f_{i}$ model a departure from kinetic equilibrium and allow the
particles to move in response to the force exerted by the bubble wall. They do not contribute
to the particle density, i.e.~$\int d^3p~\delta f_{i}=0$.
To second order in derivatives, we have to distinguish between particle and 
antiparticle perturbations, which we can expand as
\be
\mu_{i}=\mu_{i,1e}+\mu_{i,2o}+\mu_{i,2e},~~~~~~~~
\delta f_{i}=\delta f_{i,1e}+\delta f_{i,2o}+\delta f_{i,2e}.
\ee
Notice that the second order perturbations have CP-even and CP-odd
parts, which we treat separately.

Let us now concentrate on the Dirac fermion of the last section, so that we
can drop the index $i$ to simplify the notation.
We expand its distribution function to second order in derivatives as
\bear \label{expf}
f&\approx&f_{0,v_w}+f_{0,v_w}'(\gamma_w\Delta E-\mu_{1e}-\mu_{2o}-\mu_{2e})
\nonumber\\[.3cm]
&&+\frac{1}{2}f_{0,v_w}''(\gamma_w^2(\Delta E)^2-2\gamma_w\Delta E\mu_{1e}+\mu_{1e}^2)
\nonumber\\[.3cm]
&&+\delta f_{1e}+\delta f_{2o}+\delta f_{2e}.
\eear
Here $f_{0,v_w}$ denotes the equilibrium distribution (\ref{feq}) where $E$ is 
replaced by $E_0$, and  $f_{0,v_w}'=(d/dE_0)f_{0,v_w}$.
The dependence on the wall velocity is taken exact at this stage.

The evolution of $f$ is governed by the Boltzmann equation
\be
{\bf L}[f]\equiv (\dot{z}\partial_z+\dot{p}_z\partial_{p_z})f={\bf C}[f].
\ee
We look for a stationary solution, so that the explicit time derivative drops.
Plugging the ansatz (\ref{expf}) into the Boltzmann equation,
taking $\dot{z}$ and $\dot{p}_z$ from eqs.~(\ref{vk}) and (\ref{Fk}),
and subtracting the results of particles and antiparticles, we obtain for
the flow part
\bear
{\bf L}[f]|_{\rm CP-odd}&=&
-\frac{p_z}{E_0}f_{0,v_w}'\mu_2'+\gamma_wv_w\frac{(m^2)'}{2E_0}f_{0,v_w}''\mu_2
\nonumber\\[.3cm]
&&+\gamma_wv_w{\rm sign}(p_z)\frac{(m^2\theta')'}{2E_0E_{0z}}f_{0,v_w}'
\nonumber\\[.3cm]
&&+\gamma_wv_w{\rm sign}(p_z)\frac{\theta'm^2(m^2)'}{4E_0^2E_{0z}}
\left(\gamma_wf_{0,v_w}''-\frac{f_{0,v_w}'}{E_0}\right)
\nonumber\\[.3cm]
&&+\frac{\theta'm^2|p_z|}{2E_0^2E_{0z}}
\left(\gamma_wf_{0,v_w}''-\frac{f_{0,v_w}'}{E_0}\right)\mu_1'
\nonumber\\[.3cm]
&&-\gamma_wv_w{\rm sign}(p_z)\frac{(m^2\theta')'}{2E_0E_{0z}}f_{0,v_w}''\mu_1
\nonumber\\[.3cm]
&&-\gamma_wv_w{\rm sign}(p_z)\frac{\theta'm^2(m^2)'}{4E_0^2E_{0z}}
\left(\gamma_wf_{0,v_w}'''-\frac{f_{0,v_w}''}{E_0}\right)\mu_1
\nonumber\\[.3cm]
&&+\frac{p_z}{E_0}\partial_z\delta f_2-\frac{(m^2)'}{2E_0}\partial_{p_z}\delta f_2
\nonumber\\[.3cm]
&&+\frac{\theta'm^2|p_z|}{2E_0^3E_{0z}}\partial_z\delta f_1
+{\rm sign}(p_z)\left[\frac{(m^2\theta')'}{2E_0E_{0z}}
-\frac{\theta'm^2(m^2)'}{4E_0^3E_{0z}}\right]\partial_{p_z}\delta f_1.
\eear
Note that the second order perturbations present differences for particles
and antiparticles, i.e.~$\mu_{2}=\mu_{2o}-\bar\mu_{2o}$, the same as for $\delta f_2$.
The CP-even parts drop. For the first order perturbations we take 
$\mu_{1}=\mu_{1e}+\bar\mu_{1e}$, etc. 

We average the Boltzmann equation over momentum,weighting it by
1 and $p_z/E_0$. We also expand in the wall velocity, keeping only the
linear order, i.e.~$f_{0,v_w}\approx f_{0}+v_wp_zf_0'$. We then obtain
\bear \label{2}
v_w K_1 \mu_2' + v_w K_2 (m^2)' \mu_2 + u_2'  -\left\langle {\bf C}[f]\right\rangle
&=&S_{\mu}
\nonumber\\[.3cm]
-K_4 \mu_2' + v_w\tilde K_5 u_2' + v_w\tilde K_6 (m^2)'u_2 
-\left\langle \frac{p_z}{E_0}{\bf C}[f]\right\rangle
&=&S_{\theta}+S_{u}
\eear
with the source terms
\bear \label{source}
S_{\mu}&=&K_7\theta'm^2 \mu_1'
\nonumber\\
S_{\theta}&=&-v_wK_8(m^2\theta')'+v_wK_9 \theta'm^2(m^2)'
\nonumber\\
S_{u}&=&-\tilde K_{10}m^2\theta'u_1'.
\eear
The primes again denote  derivatives with respect to $z$.
The sources $S_{\mu,u}$ are related to the first order perturbations.
Notice that these are first order in $v_w$. Formally,
$S_{\mu,u}$ are one order higher in gradients than $S_{\theta}$. It will
turn out that they indeed contribute only a small fraction to the total
source term.
After momentum
integration we normalize the resulting equations by the average of the
massless Fermi--Dirac distribution
\be
\langle X\rangle=\frac{\int d^3p~X(p)}{\int d^3p f'_{0+}(m=0)}.
\ee
This normalization we also use for bosons to keep the interaction rates
for fermions and bosons equal. The plasma velocity we define as
\be \label{u}
u_2=\left\langle\frac{p_z}{E_0}\delta f_2\right\rangle.
\ee
The thermal averages read
\bear \label{av}
K_1&=&-\left\langle \frac{p_z^2}{E_0}f_0''\right\rangle,~~~~~~~~~~
\tilde K_6=\left[\frac{E_0^2-p_z^2}{2E_0^3}f_{0}'\right],
\nonumber\\[.3cm]
K_2&=&\left\langle\frac{f''_0}{2E_0}\right\rangle,~~~~~~~~~~~~~~
K_7=\left\langle \frac{|p_z|}{2E_0^2E_{0z}}\left(\frac{f_0'}{E_0}-f_0''\right)\right\rangle,
\nonumber\\[.3cm]
K_3&=&\left\langle\frac{f'_0}{2E_0}\right\rangle,~~~~~~~~~~~~~~
K_8=\left\langle \frac{|p_z|f_0'}{2E_0^2E_{0z}}\right\rangle,
\nonumber\\[.3cm]
K_4&=&\left\langle \frac{p_z^2}{E_0^2}f'_{0}\right\rangle,~~~~~~~~~~~~~
K_9=\left\langle \frac{|p_z|}{4E_0^3E_{0z}}\left(\frac{f_0'}{E_0}-f_0''\right)\right\rangle,
\nonumber\\[.3cm]
\tilde K_5&=&\left[\frac{p_z^2}{E}f_{0}'\right],~~~~~~~~~~~~~~~
\tilde K_{10}=\left[\frac{|p_z|f_0}{2E_0^3E_{0z}}\right].
\eear
The averages $\tilde K_i$ are related to averages involving $\delta f_2$. Since
we do not know the momentum dependence of  $\delta f_2$, we make the
additional assumption that these averages factorize and then use eq.~(\ref{u}), 
e.g.~$\langle p_z^3 \delta f_2\rangle\approx[p_z^2E_0f_{0,v_w}]u$.
We normalize these averages by the massive distribution of the boson or
fermion under consideration, i.e.~$\int d^3p~f_{0,v_w}$.
Since there is some arbitrariness in this procedure, we will test the impact
of these averages, which turns out to be small\footnote{Depending
on how we precisely treat the averages involving $\delta f_2$, there can
also arise a source term of the form $(m^2)'\theta'u_1$. We do not discuss it
in more detail since the source terms related to the first order perturbations 
are small anyway.}.

The collision integrals read \cite{CJK00}
\bear
\left\langle {\bf C}[f]\right\rangle&=&\Gamma^{\rm inel} \sum \mu_{i,2}
\nonumber\\[.3cm] 
\left\langle \frac{p_z}{E_0}{\bf C}[f]\right\rangle&=&-\Gamma^{\rm tot}u_2,
\eear
where $\Gamma^{\rm inel}$ and $\Gamma^{\rm tot}$ are the inelastic and
total interaction rates, respectively. The negative sign in front of $\Gamma^{\rm tot}$
is related to our sign convention for the plasma velocity (\ref{u}).

The transport equations of the first order perturbations look very similar to eq.~(\ref{2})
\bear \label{1}
v_w K_1 \mu_1' + v_w K_2 (m^2)' \mu_1 + u_1'  -\Gamma^{\rm inel} \sum \mu_{i,1}
&=&v_wK_{3}(m^2)'
\nonumber\\[.3cm]
-K_4 \mu_1' + v_w\tilde K_5 u_1' + v_w\tilde K_6 (m^2)'u_1+\Gamma^{\rm tot}u_1
&=&0.
\eear
The source term is now first order in derivatives and CP-even. Note that here
also the quite large annihilation rates enter in  $\Gamma^{\rm inel}$.

In eqs.~(\ref{2}) and (\ref{1}) we can approximately eliminate the plasma velocity
to obtain diffusion equations for chemical potentials. From the coefficient of
the $\mu''$ term we can read off the diffusion constant as \cite{CJK00}
\be\label{D}
D=\frac{K_4}{K_1\Gamma^{\rm tot}}.
\ee

Our source terms (\ref{source}) agree with those obtained from the Schwinger--Keldysh
formalism. However, in ref.~\cite{PSW1} there is one extra source term, related to the 
gradient
renormalization of the Wigner function. This term seems to be missing in the Dirac
equation approach. It is of order $m^4$, like the $K_9$-part of $S_{\theta}$. We will
demonstrate in the next section that these terms are subleading.

\section{Top transport: an example}
We now apply the general results (\ref{2}) and (\ref{1}) to top transport in an effective 
SM with dimension-6 operators \cite{Z93,ZLWY94,GSW04,HO04,BFHS04}. 
The model contains a single Higgs doublet, whose potential is stabilized by a $\phi^6$
interaction
\be
V(\phi)=-{\mu^2 \over 2}\phi^2+{\lambda \over 4}\phi^4+{1\over{8M^2}}\phi^6.
\ee
This potential has two free parameters, the suppression scale $M$ of the 
dimension-6 operator and the quartic coupling $\lambda$.  The latter can be eliminated
in terms of the physical Higgs mass $m_H$. Since the potential is stabilized by the
$\phi^6$ term, $\lambda$ can be negative. In this case a barrier in Higgs potential
is present at tree-level, which triggers a first order electroweak phase transition.
Computing the 1-loop thermal potential, it was shown in ref.~\cite{BFHS04}
that the phase transition is strong enough to avoid baryon number washout,
i.e.~$\xi=\langle \phi\rangle_{T_c}/T_c>1.1$ \cite{M98}, if $M\lsim850$ GeV and  
$m_H=115$ GeV. Taking $M=500$ GeV, a strong phase transition is present for
$m_H\lsim180 $ GeV. Thus the model allows for a strong phase transition in a 
large part of its parameter space. In ref.~\cite{BFHS04} also the wall thickness
has been determined, showing that $3\lsim L_wT_c\lsim 16$. The gradient expansion
discussed in section 2 is therefore justified in almost the full parameter space.
The thinnest walls correspond to a very strong phase transition, $\xi\sim3$,
where the model is close to metastability of the symmetric phase. In the following
we will approximate the wall profile by a hyperbolic tangent,
$\phi(z)=(v_c/2)(1-\mbox{tanh}(z/L_w))$.

Dimension-6 operators also induce new sources of CP-violation. In addition to the
ordinary Yukawa interaction of the top quark, $y_t\Phi t^cq_3$, we have an operator
$(x_t/M^2)(\Phi^{\dagger}\Phi)\Phi t^cq_3$ \cite{ZLWY94}. We denote the relative
phase between the two couplings as $\varphi_t={\rm arg}(yx^*)$. Then the top
develops a position dependent complex phase $\theta_t$ along the bubble wall $\phi(z)$,
with
\begin{equation} \label{theta}
\tan\theta_t(z)\approx \sin\varphi_t\frac{\phi^2(z)}{2M^2}\left|\frac{x_t}{y_t}\right|.
\end{equation} 
So all necessary ingredients are present to apply the formalism discussed in 
the previous sections. 

For the generation of the baryon asymmetry, the most important particle species are 
the left- and right-handed top quarks, and the Higgs bosons. We will show that 
the latter have only a minor impact. We ignore leptons, which are only produced by
small Yukawa couplings. 
In contrast to all previous investigations we include the $W$ scatterings with a finite rate 
$\Gamma_W$.  This procedure allows us to study the perturbations of bottom and top
quarks separately. The top quark source is no longer locked to the
bottom degrees of freedom, which would lead to a larger or smaller baryon asymmetry,
depending on the wall velocity. 
The other interactions we
take into account are the top Yukawa interaction, $\Gamma_y$, the weak and 
strong sphalerons, $\Gamma_{ws}$ and $\Gamma_{ss}$, the top helicity flips, 
$\Gamma_m$, and Higgs number violation $\Gamma_h$.
The latter two are only present in the broken phase.

In a first step we compute the left-handed quark density, assuming that baryon 
number is conserved. Later on, the left-handed quark density will be converted
into a baryon asymmetry by the weak sphalerons. The transport equations for
chemical potentials of left-handed SU(2) doublet tops $\mu_{t,2}$,  left-handed 
SU(2) doublet bottoms $\mu_{b,2}$, left-handed SU(2) singlet tops $\mu_{t^c,2}$, 
Higgs bosons $\mu_{h,2}$, and the corresponding
plasma velocities read 
\bear \label{mus}
3v_wK_{1,t}\mu_{t,2}'+3v_wK_{2,t}(m_t^2)'\mu_{t,2}+3u_{t,2}'&&
\nonumber\\
-3\Gamma_y(\mu_{t,2}+\mu_{t^c,2}+\mu_{h,2})-6\Gamma_m(\mu_{t,2}+\mu_{t^c,2})
-3\Gamma_W(\mu_{t,2}-\mu_{b,2})&&
\nonumber\\
-3\Gamma_{ss}[(1+9K_{1,t})\mu_{t,2}+(1+9K_{1,b})\mu_{b,2}+(1-9K_{1,t})\mu_{t^c,2}]
&=&3K_{7,t}\theta_t'm_t^2\mu_{t,1}'
\nonumber\\[.5cm]
3v_wK_{1,b}\mu_{b,2}'+3u_{b,2}'&&
\nonumber\\
-3\Gamma_y(\mu_{b,2}+\mu_{t^c,2}+\mu_{h,2})
-3\Gamma_W(\mu_{b,2}-\mu_{t,2})&&
\nonumber\\
-3\Gamma_{ss}[(1+9K_{1,t})\mu_{t,2}+(1+9K_{1,b})\mu_{b,2}+(1-9K_{1,t})\mu_{t^c,2}]
&=&0
\nonumber\\[.5cm]
3v_wK_{1,t}\mu_{t^c,2}'+3v_wK_{2,t}(m_t^2)'\mu_{t^c,2}+3u_{t^c,2}'&&
\nonumber\\
-3\Gamma_y(\mu_{t,2}+\mu_{b,2}+2\mu_{t^c,2}+2\mu_{h,2})-6\Gamma_m(\mu_{t,2}+\mu_{t^c,2})&&
\nonumber\\
-3\Gamma_{ss}[(1+9K_{1,t})\mu_{t,2}+(1+9K_{1,b})\mu_{b,2}+(1-9K_{1,t})\mu_{t^c,2}]
&=&3K_{7,t}\theta_t'm_t^2\mu_{t^c,1}'
\nonumber\\[.5cm]
2v_wK_{1,h}\mu_{h,2}'+2u_{h,2}'&&
\nonumber\\
-3\Gamma_y(\mu_{t,2}+\mu_{b,2}+2\mu_{t^c,2}+2\mu_{h,2})-2\Gamma_h\mu_{h,2}&=&0
\eear
\bear \label{us}
-3K_{4,t}\mu_{t,2}'+3v_w\tilde K_{5,t}u_{t,2}'+3v_w\tilde K_{6,t}(m_t^2)'u_{t,2}
+3\Gamma^{\rm tot}_tu_{t,2}&=&
\nonumber\\
=-3v_wK_{8,t}(m^2_t\theta_t')'+3v_wK_{9,t} \theta_t'm^2_t(m_t^2)'
-3\tilde K_{10,t}m^2_t\theta_t'u_{1,t}'&&
\nonumber\\[.5cm]
-3K_{4,b}\mu_{b,2}'+3v_w\tilde K_{5,b}u_{b,2}'+3\Gamma^{\rm tot}_bu_{b,2}&=&0
\nonumber\\[.5cm]
-3K_{4,t}\mu_{t^c,2}'+3v_w\tilde K_{5,t}u_{t^c,2}'
+3v_w\tilde K_{6,t}(m_t^2)'u_{t^c,2}+3\Gamma^{\rm tot}_tu_{t^c,2}&=&
\nonumber\\
=-3v_wK_{8,t}(m^2_t\theta_t')'+3v_wK_{9,t} \theta_t'm^2_t(m_t^2)'
-3\tilde K_{10,t}m^2_t\theta_t'u_{1,t^c}'&&
\nonumber\\[.5cm]
-2K_{4,h}\mu_{h,2}'+2v_w\tilde K_{5,h}u_{h,2}'+2\Gamma^{\rm tot}_hu_{h,2}&=&0
\eear
In eqs.~(\ref{us}) $\Gamma_W$ can be neglected since the plasma velocities of $t$ and
$b$ are damped by the much faster gluon scatterings. We have used baryon number 
conservation to express the sphaleron interaction in terms of $\mu_{t,2}$,
$\mu_{b,2}$ and $\mu_{t^c,2}$ \cite{HN95}. A possible source term for the bottom
quark is suppressed by $(m_b/m_t)$ and therefore neglected.

The first order perturbations of $t$ can be computed from
\bear
3v_wK_{1,t}\mu_{t,1}'+3v_wK_{2,t}(m_t^2)'\mu_{t,1}+3u_{t,1}'
-3\Gamma^{\rm tot}_t\mu_{t,1}&=&3v_wK_{3,t}(m_t^2)'
\nonumber\\[.3cm]
-3K_{4,t}\mu_{t,1}'+3v_w\tilde K_{5,t}u_{t,1}'+3v_w\tilde K_{6,t}(m_t^2)'u_{t,1}
+3\Gamma^{\rm tot}_tu_{t,1}&=&0.~~
\eear
The damping of $u_{t,1}$ is dominated by gluon annihilation, the rate of which
we have approximated by $\Gamma^{\rm tot}_t$. Other scatterings have
been neglected. To this approximation the chemical potentials of $t$ and
$t^c$ are identical. This guarantees that no direct source for baryon number
is induced. Such a source can be generated if $\mu_{t,1}\neq\mu_{t^c,1}$.
It leads to spurious effects in the baryon asymmetry.
Its appearance shows that an inconsistent approximation pattern has been used.

We can now compute the chemical potential of left-handed quarks, 
$\mu_{B_L}=\mu_{q_1,2}+ \mu_{q_2,2}+(\mu_{t,2}+\mu_{b,2})/2$. Assuming
again baryon number conservation, we obtain 
\begin{equation}
\mu_{B_L}=\frac{1}{2}(1+4K_{1,t})\mu_{t,2}+\frac{1}{2}(1+4K_{1,b})\mu_{b,2}
-2K_{1,t}\mu_{t^c,2}.
\end{equation}
The baryon asymmetry is then given by \cite{CJK00}
\begin{equation} \label{eta1}
\eta_B=\frac{n_B}{s}=\frac{405\Gamma_{ws}}{4\pi^2v_wg_*T}\int_0^{\infty}
dz ~\mu_{B_L}(z)e^{-\nu z},
\end{equation}
where is $\Gamma_{ws}$ the weak sphaleron rate and
$\nu=45\Gamma_{ws}/(4v_w)$. The effective number of
degrees of freedom in the plasma is $g_*=106.75$.
In eq.~(\ref{eta1}) the
weak sphaleron rate has been suddenly switched off in the
broken phase, $z<0$. The exponential factor in the integrand
accounts for the relaxation of the baryon number if the wall
moves very slowly.  Note that we have performed our computation
in the wall frame. Therefore, strictly speaking eq.~(\ref{eta1}) gives
the baryon asymmetry in that frame. To first order in $v_w$ this
is identical to the baryon asymmetry in the plasma frame.  

In our numerical evaluations we use the following values for the 
weak sphaleron rate \cite{Mws}, the strong sphaleron rate \cite{Mss}, the top
Yukawa rate \cite{HN95}, the top helicity flip rate, the Higgs number violating
rate \cite{HN95}, the quark diffusion constant  \cite{JPT95} and
the Higgs diffusion constant  \cite{CJK00} 
\begin{eqnarray} \label{rates}
&&\Gamma_{ws}=1.0\times10^{-6}T, \quad  \quad \Gamma_{ss}=4.9\times10^{-4}T, 
\nonumber \\
&&\Gamma_y=4.2\times10^{-3}T, \quad \quad ~~\Gamma_m=\frac{m_t^2(z,T)}{63T},
\nonumber \\
&&\Gamma_h=\frac{m_W^2(z,T)}{50T},\quad \quad ~~~~D_q=\frac{6}{T},
\nonumber \\
&&D_h=\frac{20}{T}.
\end{eqnarray}
We use eq.~(\ref{D}) to infer the total interaction rates from the diffusion constants.
In this procedure we evaluate the thermal averages at $z=0$, i.e.~in the center of the
bubble wall.
The $W$ scatterings we approximate as $\Gamma_W=\Gamma^{\rm tot}_h$. 
The bottom quark is taken as massless, and the Higgses we count as 2 massless
complex degrees of freedom.  The rates of eq.~(\ref{rates}) have been computed
in the plasma frame. We assume that, to leading order in $v_w$, they can also be
used in the wall frame. 

\begin{figure}
\begin{center}
   \epsfig{file=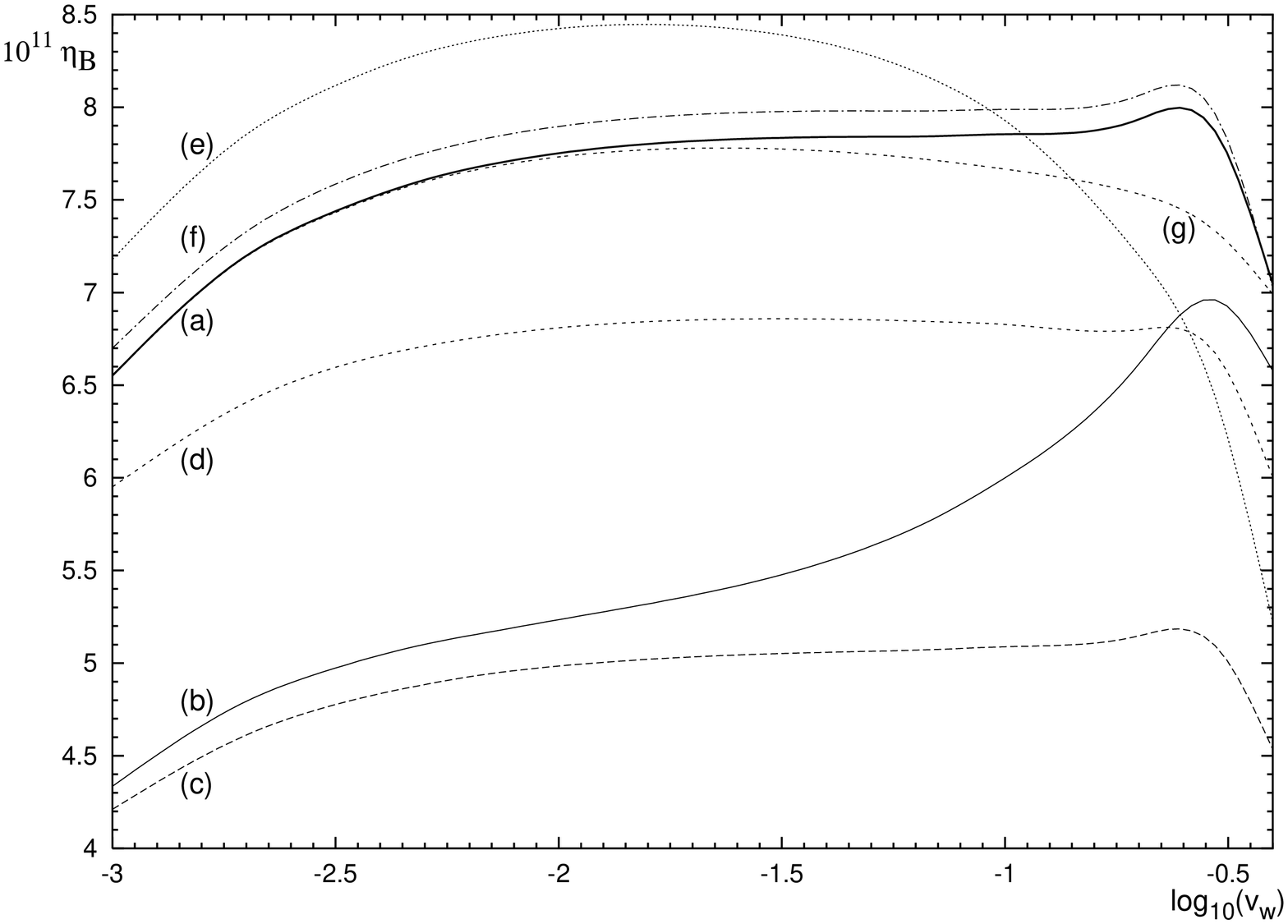,width=90mm,angle=270}
\end{center}
\caption{The baryon asymmetry as a function of $v_{w}$ for $\xi=1.5$, $M=6$
  and $L_{\rm w}=8$. The labeling is explained in the text.}
\label{figure1}
\begin{center}
\epsfig{file=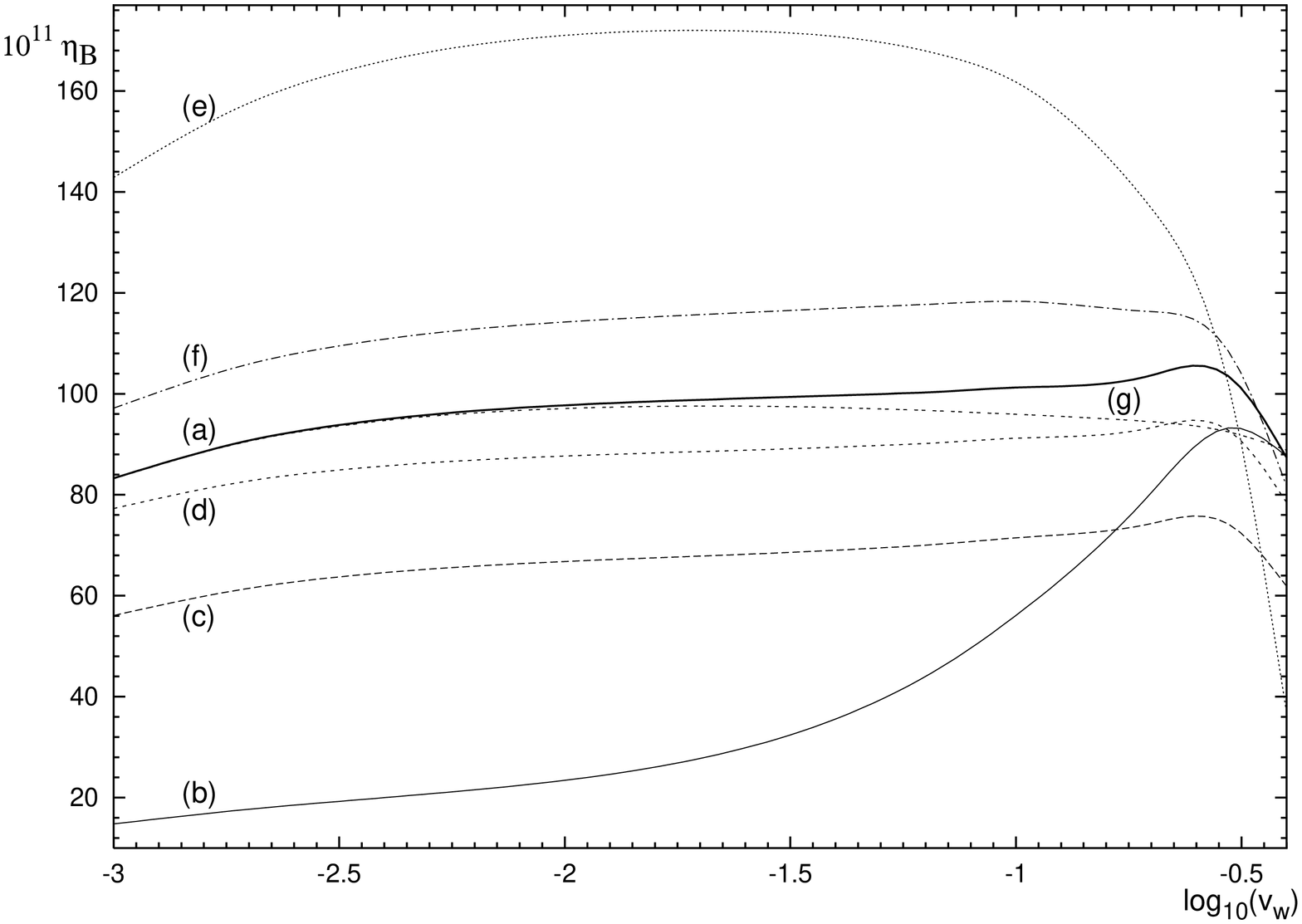,width=90mm,angle=270}
\end{center}
\caption{The baryon asymmetry as a function of $v_{w}$ for $\xi=2.5$, $M=6$
  and $L_{\rm w}=3$.}
\label{figure2}
\end{figure}

To demonstrate the relevance of the various contributions to the full transport equations, 
we compare the baryon asymmetry computed in different
approximations for two typical parameter settings. We take $|x_t|=1$ and 
maximal CP violation $\sin\varphi_t=1$.
Fig.~\ref{figure1} shows $\eta_B$ as a function of the wall velocity
$v_w$. The other parameters we have chosen as $\xi=1.5$, $M=6$ and $L_{w}=8$. 
These values correspond to a setting where the baryon asymmetry is
close to the observed value 
$\eta_B=(8.9\pm0.4)\times10^{-11}$ \cite{CMB,LSS}.
In Fig.~\ref{figure2} we use $\xi=2.5$, $M=6$ and $L_{w}=3$, i.e.~a 
very strong phase transition with a small wall width. In both
figures the bold solid line (a) indicates $\eta_B$ using the source terms $S_{\theta}$
and keeping the full $z$-dependence of the thermal averages (\ref{av}). 

In (b) we drop the space-dependence of the thermal averages. We rather evaluate them
at the center of the bubble wall, i.e.~$K_{i,t}(z)\equiv K_{i,t}(z=0)$.  Formally, 
the space-dependence of the thermal averages is a higher order effect in gradients.
But this approximation considerably underestimates the baryon asymmetry, especially for 
small wall velocities and thin bubble walls. The full  $z$-dependence reduces the
impact of the wall velocity on $\eta_B$.

The long-dashed line (c) shows the result when we resubstitute $E_{0z}\rightarrow E_0$,
going back to the dispersion relation of ref.~\cite{CJK00}. This would considerably
reduce the baryon asymmetry, in particular for weaker phase transitions (Fig.~\ref{figure1}).

Neglecting the Higgs bosons in the transport eqs.~(\ref{mus}) and (\ref{us}) leads 
to a reduction of $\eta_B$ by $\simeq 10\%$ (d), almost independent of the
wall velocity and the strength of the phase transition.

Taking the $W$ scatterings to equilibrium (e) has a substantial effect on
the resulting baryon asymmetry, especially for strong phase transitions.
In Fig.~\ref{figure2} it overestimates $\eta_B$ by a factor of almost 2 for 
$v_w<0.1$. For large wall velocities there is an underestimate of $\eta_B$
by a similar size. Keeping W scatterings finite results in a much milder
$v_w$-dependence of the baryon asymmetry.

The dash-dotted line (f) adds the contributions of $S_{\mu}+S_u$ to line (a). 
The effect of these source terms is quite small, consistent with the fact that
they are of higher order in gradients.
They enhance the baryon asymmetry in the whole $v_{w}$-range only by a 
few percent. 

Line (g) shows the effect of switching off  the terms proportional to
$\tilde{K}_5$ and $\tilde{K}_6$. If these terms are neglected, the 
final result is reduced by a contribution proportional to the wall velocity. 
It demonstrates that the precise treatment
of the averages involving $\delta f$ has only a minor impact on the
baryon asymmetry\footnote{Numerically there is also not much difference 
to the prescription used, for instance, in ref.~\cite{BFHS04}, where
plasma velocities were included in the fluid ansatz, rather than using
a general $\delta f$. Then, for example, the $u_2'$ term in eq.~(\ref{2}) obtains
an additional coefficient $\sim 1.1$.}.

Altogether the examples demonstrate that the leading 
contribution to $\eta_B$ comes from the source $S_{\theta}$. 
The baryon asymmetry gets considerably enhanced by using the
dispersions relation with the correct factors of $E_{0z}$ and keeping
the space-dependence of the thermal averages. The finite $W$ scattering
rate has a sizable effect, the direction of which depends on the wall
velocity. The resulting $v_w$-dependence of the baryon asymmetry
is rather mild. The baryon asymmetry grows slowly with increasing 
$v_{\rm w}$ and reaches a maximum at $v_{w}\simeq$ 0.2--0.3. 
Taking the Higgs bosons or the $S_{\mu}+S_u$  sources
into account is less important. Their effect is not larger than typical
uncertainties from higher order terms in the gradient expansion.

The source $S_{\theta}$ consists of two parts, proportional to $K_8$
and $K_9$. The latter has an additional factor $m^2$, leading to an extra 
suppression, in particular for weak phase transitions. For instance,
taking $v_w=0.1$ and the parameter set of Fig.~\ref{figure1}, the
$K_9$-part contributes only about 15\% to the total baryon asymmetry.
As indicated earlier, there is an extra source term in ref.~\cite{PSW1}, 
which is related to the gradient renormalization of the Wigner function.
It also has an extra factor of $m^2$ and therefore should also be
sub-leading in our case.

Of course, the baryon asymmetry also depends on the precise values of 
the interaction rates (\ref{rates}). For instance, reducing the quark diffusion 
constant by 10\% leads to an about 7\% reduction in the baryon asymmetry
(taking $v_w=0.1$ and the  parameter set of Figure \ref{figure1}). 
Changing $\Gamma_W$ by 10\% affects $\eta_B$ to less than 1\%, even
for   $v_w\sim0.01$, where the impact of the $W$ scatterings is particularly
large. 

\begin{figure}
\begin{center}
   \epsfig{file=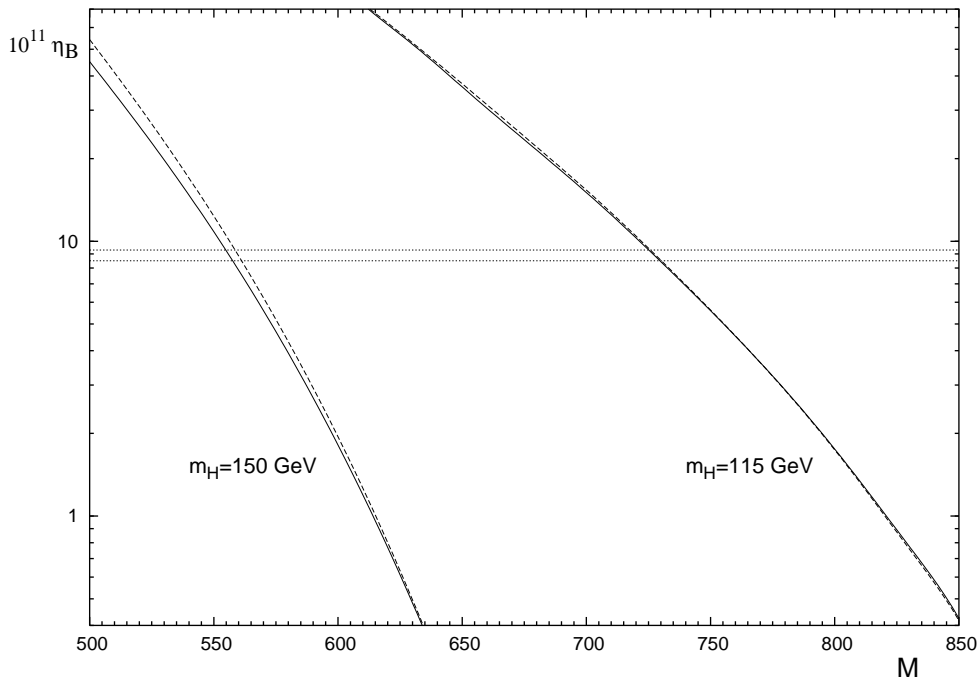,width=90mm,angle=270}
\end{center}
\caption{The baryon asymmetry in the SM with low cut-off for two different Higgs
  masses as a function
of $M$ (in units of GeV) for $v_w=0.01$ (solid) and $v_w=0.3$ (dashed). The horizontal lines indicate the error band of the observed value.}
\label{model}
\end{figure}

Figure \ref{model} displays the baryon asymmetry in the SM with a low cut-off as
a function of the cut-off scale $M$. We consider two different Higgs masses $m_H=115$ GeV and $m_H=150$
GeV and two wall velocities $v_w=0.01$ and 0.3. For each value
of $M$ the corresponding strength of the phase transition and bubble width are
computed as in ref.~\cite{BFHS04}. As expected $\eta_B$ increases rapidly with 
decreasing cut-off scale $M$. The asymmetry has only a minor dependence on the wall velocity.
In both cases it is possible to generate the measured
baryon asymmetry for a reasonably small value of $M$.

\section{Conclusions}
We have studied aspects of top transport in electroweak baryogenesis. 
We have computed its CP-violating source term in the WKB approximation, using
the one-particle Dirac equation in the wall background.
When the top dispersion relation is correctly boosted to a general Lorentz frame,
the Schwinger-Keldysh result \cite{PSW,PSW1} for the semiclassical force 
term is obtained in eq.~(\ref{Fk}). The CP-violating source term is enhanced
with respect to ref.~\cite{CJK00}. We have only considered the case of a single
Dirac fermion, but our results should simply generalize to mixing fermions, such
as the charginos in the MSSM.

In our computation we cannot obtain the extra source term of ref.~\cite{PSW1}, 
which is related to the gradient renormalization of the Wigner function. In the
case of top transport this term is subleading since it is of order $m^4$. In our approach,
of course, we also cannot obtain source terms related to quantum mechanical
oscillations between different fermion flavors. In the case of the top quark this
effect is obviously not present, but it can be relevant for the charginos in the MSSM
\cite{thomas05}.

We have demonstrated the numerical significance of the corrected dispersion relations
in the SM augmented by dimension-6 operators. This effect alone enhances the 
baryon asymmetry by a factor of up to about 2.
We have also improved on the transport equations, keeping scatterings with $W$
bosons at a finite rate. Depending on the wall velocity and the wall thickness, putting 
the $W$ scatterings to equilibrium (as was done so far in the literature) can increase
or decrease the baryon asymmetry by a factor of 2. It would be interesting to study
the impact of this effect in supersymmetric models, where the SU(2) supergauge 
interactions have been put to equilibrium as well.   

We have shown that the position dependence of
the thermal averages in the transport equations has a substantial impact on the
baryon asymmetry, even though it is formally a higher order effect in the gradient
expansion.
Finally, the influence of the Higgs bosons on transport turned out to be small, as is
the contribution of the sources $S_{\mu,u}$ (\ref{source}).
In total, depending on the model parameters, 
our refinements can increase the baryon asymmetry by a factor of up to about 5.

The rather large impact of the precise treatment of the $W$ scattering rate and
the space-dependence of the thermal averages probably indicate that there
is still a substantial uncertainty related to transport. 

In a forthcoming publication we will apply the framework presented here to compute
the baryon asymmetry in the two Higgs doublet model \cite{FHS}

\section*{Acknowledgements}
We thank D.~B\"odeker, M.~Seniuch and S.~Weinstock 
for valuable discussions. The work of L.F.~was supported by the DFG,
grand FOR 339/2-1.


\end{document}